# A SIMPLE PROOF OF KORAY'S THEOREM

HÉCTOR HERMIDA-RIVERA[1] • h.hermidarivera@hse.ru November 21, 2026

*In this note, I provide a direct and self-contained proof of Koray's (2000) theorem, by which a neutral and unanimous social choice function is universally self-selective if and only if it is dictatorial. Koray's (2000) original proof relies on results by Houthakker (1950) and Sen (1970) to construct a social welfare function for each alternative set. He then shows that these functions are Paretian and independent of irrelevant alternatives, allowing him to apply Arrow's (1950, 1951) impossibility theorem. Instead, I construct a cyclic preference profile and show that the voter who ranks the chosen alternative first is the dictator on every finite nonempty alternative set.*

## 1. INTRODUCTION

**Overview**. Koray (2000) studied the ubiquitous problem of choosing a social choice function. In doing so, he introduced the notion of universal self-selectivity (p. 985), by which a neutral social choice function must always be able to select itself from any finite set of neutral social choice functions it belongs to. He then showed that a neutral and unanimous social choice function is universally self-selective if and only if it is dictatorial (Koray, 2000, Theorem 2, p. 990).

Koray's (2000) proof strategy is complex. In Theorem 1 (pp. 987–988), he derives Pareto efficiency and independence of irrelevant alternatives[2] from unanimity, neutrality, and universal self-selectivity. Then, Koray (2000, Proposition 1, p. 989) draws on results by Sen (1970) and Houthakker (1950) to construct—for each alternative set—a social welfare function from the original social choice function. Further, he shows that the induced social welfare functions are Paretian and independent of irrelevant alternatives (Proposition 2, p. 990), so Arrow's (1950, 1951) impossibility theorem yields a dictator for each set of at least three alternatives. Finally, he shows that the same voter dictates the original social choice function on every finite nonempty alternative set (Theorem 2, pp. 990–991).

**Contribution**. In this paper, I provide a simple and self-contained proof of Koray's (2000) result. In Claim 2.1, I derive independence of irrelevant alternatives just as he does, and in Claim 2.2, I derive Pareto efficiency using the slightly simpler argument of Hermida-Rivera & Kerman (2025, Claim 3.2, p. 6). From then on, my proof is completely different. In Claim 2.3, I construct a cyclic preference profile to identify a voter whose preferred alternative is chosen whenever all other voters favor the other alternative in a binary choice. And in Claim 2.4, I use this binary choice property together with Pareto efficiency and independence of irrelevant alternatives to show that the same voter's top-ranked alternative is chosen at every profile.

**Literature**. Among the multiple papers building on Koray's (2000) contribution, only Koray & Slinko (2008, Theorem 2 and Corollary 1, p. 141) provide an alternative proof of his theorem. Their result applies to societies with at least three voters, and their proof identifies a decisive voter by successively splitting decisive coalitions (Lemmas 5–6, pp. 145–146). Although my proof differs substantially from theirs, the techniques I use do have antecedents in the literature. In their analysis of neutral strategyproof social choice functions, Svensson & Reffgen (2014, Proof of Theorem 3, p. 14) use the same cyclic construction used in Claim 2.3. And in the proof of Claim 2.4, I adapt a three-alternative argument that Basu Mallik & Bhattacharya (2026, Lemma 3, p. 9) use to prove that Pareto efficiency and contraction consistency imply dictatorship on nonempty subsets of a fixed finite set with at least three alternatives (Theorem 2, p. 3).

[1] *Faculty of Economic Sciences, HSE University, 11 Pokrovsky Bulvar, Moscow, 109028, Russia.*

[2] For social choice functions, Koray's (2000) independence of irrelevant alternatives and Basu Mallik & Bhattacharya's (2026, Definition 3, p. 3) contraction consistency require the chosen alternative to remain chosen whenever it belongs to the restricted alternative set. However, for social welfare functions, independence of irrelevant alternatives requires the social ranking of any pair to depend only on how voters rank that pair.

**Notation**. Throughout the paper, I adopt the following notational conventions: (**1**) for any collection of individual objects $(\vartheta_i)_{i\in N}$, the corresponding profile is denoted by the unsubscripted letter $\vartheta = (\vartheta_i)_{i\in N}$; (**2**) unless otherwise specified, accented symbols denote objects of the same kind as their unaccented counterparts; (**3**) variable ranges in quantifiers and set-builder notation are omitted when clear from the context.

**Structure**. Section 2 sets out the environment, Section 3 defines dictatorships, Section 4 introduces the base axioms, Section 5 formalizes universal self-selectivity, and Section 6 provides a simple proof of Koray's (2000) result.

## 2. ENVIRONMENT

The *environment* is a 4-tuple $(N, I_m, \mathcal{L}(I_m)^N, F)$. Let $N = \{1, \dots, n\}$ be a *finite nonempty set of voters* $i, j, k, \dots$. Given any $m \in \mathbb{N} = \{1, 2, \dots\}$, let $I_m = \{1, \dots, m\}$ be a *finite set of alternatives*, and let

$$\mathcal{L}(I_m)^N = \{(R_i)_{i\in N} \mid (\forall i)[R_i \text{ is a linear order on } I_m]\} \tag{1}$$

be the *set of all linear order profiles* $R$ *on the alternative set*.[3] Finally, let

$$F : \bigcup_{m\in\mathbb{N}} \mathcal{L}(I_m)^N \to \mathbb{N} \tag{2}$$

be any *social choice function* such that $F(R) \in I_m$ for all $m$ and $R$.

Let $\mathcal{F} = \{F \mid F \text{ is a social choice function}\}$.

## 3. DICTATORSHIPS

Given any $m$, $R$, $i$, and $l \in \{1, \dots, m\}$, let $t_l(R_i)$ be the *alternative in the* $l^{th}$ *position at* $R_i$.

DEFINITION 1 — DICTATORSHIP. Given any $i$, *voter* $i$*'s dictatorship* $d_i \in \mathcal{F}$ always chooses voter $i$'s top-ranked alternative. Formally,

$$(\forall m, R)[d_i(R) = t_1(R_i)] \tag{3}$$

Let $\mathcal{F}_d = \{d_i \mid i \in N\}$.

## 4. AXIOMS

AXIOM 1 — UNANIMITY. $F$ is *unanimous* if and only if it chooses any alternative top-ranked by every voter. Formally,

$$(\forall m, R, x)[((\forall i)[t_1(R_i) = x]) \Rightarrow (F(R) = x)] \tag{4}$$

Let $\mathcal{F}_u = \{F \in \mathcal{F} \mid F \text{ is unanimous}\}$.

Given any $m$, let $\sigma : I_m \to I_m$ be any *permutation of the alternative set*. Given any $m$, $R$, and $\sigma$, let the *linear order profile* $\sigma R$ satisfy

$$(\forall i, x, y)[(x(\sigma R)_i y) \Leftrightarrow (\sigma(x) R_i \sigma(y))] \tag{5}$$

[3] A *linear order* is a *complete*, *transitive*, and *antisymmetric* binary relation. Thus, unlike a *strict linear order*, it is also *reflexive*.

AXIOM 2 — NEUTRALITY. $F$ is *neutral* if and only if it is independent of the labels of the alternatives. Formally,

$$(\forall m, R, \sigma)[\sigma(F(\sigma R)) = F(R)] \tag{6}$$

Let $\mathcal{F}_n = \{F \in \mathcal{F} \mid F \text{ is neutral}\}$.

The domain of any $F \in \mathcal{F}_n$ can be extended to linear order profiles on any finite nonempty set. Given any $m$, let $A$ be any *finite nonempty set of elements* $a, b, \ldots$ with cardinality $m$, let $L \in \mathcal{L}(A)^N$ be any *linear order profile on* $A$, and let $\mu : I_m \to A$ be any *bijection*. Given any $L$ and $\mu$, let the *linear order profile* $\mu L \in \mathcal{L}(I_m)^N$ satisfy

$$(\forall i, x, y)[(x(\mu L)_i y) \Leftrightarrow (\mu(x) L_i \mu(y))] \tag{7}$$

Then, given any $F \in \mathcal{F}_n$ and any two bijections $\mu, \nu : I_m \to A$, neutrality implies that

$$(\forall L)[\mu(F(\mu L)) = \nu(F(\nu L))] \tag{8}$$

Hence, given any $\mu$ and $F \in \mathcal{F}_n$, extend $F$ by setting

$$(\forall L)[F(L) = \mu(F(\mu L))] \tag{9}$$

Similarly, given any $\mu$ and $l \in \{1, \ldots, m\}$, set

$$(\forall i, L)[t_l(L_i) = \mu(t_l((\mu L)_i))] \tag{10}$$

## 5. UNIVERSAL SELF-SELECTIVITY

In order to introduce Koray's (2000, p. 985) universal self-selectivity, I first recall how a linear order profile on alternatives induces a weak order profile on any finite set of neutral social choice functions through the alternatives they choose. Let $\mathcal{B} \subsetneq \mathcal{F}_n$ be any *finite nonempty set of neutral social choice functions* $G, H, \ldots$. Given any $m$, $R$, and $\mathcal{B}$, let the *weak order profile* $L^R$ *on* $\mathcal{B}$ satisfy

$$(\forall i, G, H)[(G L^R_i H) \Leftrightarrow (G(R) R_i H(R))] \tag{11}$$

and let $L$ be any *linear order profile on* $\mathcal{B}$ satisfying $L_i \subseteq L^R_i$ for all $i$. Then, any such $L$ is a *linear order profile compatible with* $L^R$. Given any $F \in \mathcal{F}_n$, let $\mathcal{A} \subsetneq \mathcal{F}_n$ be any *finite set satisfying* $F \in \mathcal{A}$.

AXIOM 3 — UNIVERSAL SELF-SELECTIVITY. $F \in \mathcal{F}_n$ is *universally self-selective* if and only if, for every linear order profile on alternatives and every finite set of neutral social choice functions containing $F$, there exists a compatible linear order profile on that set at which $F$ chooses itself. Formally,

$$(\forall m, R, \mathcal{A})(\exists L)[F(L) = F] \tag{12}$$

Let $\mathcal{F}_s = \{F \in \mathcal{F}_n \mid F \text{ is universally self-selective}\}$.

## 6. RESULT

THEOREM 1 (Koray, 2000). *A neutral and unanimous social choice function is universally self-selective if and only if it is dictatorial. Formally,*

$$(\forall F \in \mathcal{F}_n \cap \mathcal{F}_u)[(F \in \mathcal{F}_s) \Leftrightarrow (F \in \mathcal{F}_d)] \tag{13}$$

PROOF OF THEOREM 1. Consider any $F \in \mathcal{F}_n \cap \mathcal{F}_u$. Then, there are two statements to show:

1. $(F \in \mathcal{F}_d) \Rightarrow (F \in \mathcal{F}_s)$,
2. $(F \in \mathcal{F}_s) \Rightarrow (F \in \mathcal{F}_d)$.

STATEMENT 1. $(F \in \mathcal{F}_d) \Rightarrow (F \in \mathcal{F}_s)$.

The proof is direct. Let $F = d_i$ for some $i$, and consider any $m$, $R$, and $\mathcal{A}$. Since $d_i(R) = t_1(R_i)$, it follows that $d_i L_i^R G$ for all $G \in \mathcal{A}$. Hence, there exists some $L$ satisfying $t_1(L_i) = d_i$. Fix any such $L$. Then, $d_i(L) = t_1(L_i) = d_i$. Thus, $F \in \mathcal{F}_s$.

STATEMENT 2. $(F \in \mathcal{F}_s) \Rightarrow (F \in \mathcal{F}_d)$.

Let $F \in \mathcal{F}_s$. If $n = 1$, unanimity implies that $F = d_1$. So, let $n \geqslant 2$. Given any $m$, let $B$ be any *nonempty subset of* $I_m$. The proof follows from Claims 2.1 to 2.4.

CLAIM 2.1. $(\forall m, R, B)[(F(R) \in B) \Rightarrow (F(R|_B) = F(R))]$.

The proof is direct. Given any $l \in \mathbb{N}$, let $q_l \in \mathcal{F}$ satisfy

$$(\forall m, R)[q_l(R) = t_{\min\{l,m\}}(R_1)] \tag{14}$$

Then, $q_l \in \mathcal{F}_n$ for all $l \in \mathbb{N}$, and $\{q_l(R) \mid l \in I_m\} = I_m$ for all $m$ and $R$. Consider any $m$, $R$, and $B$ such that $F(R) \in B$, and set $x = F(R)$. For each $y \in B \setminus \{x\}$, choose some $G_y \in \mathcal{F}_n$ satisfying $G_y(R) = y$. Set $G_x = F$, and let $\mathcal{A} = \{G_y \mid y \in B\}$. By universal self-selectivity, there exists some $L$ satisfying $F(L) = F$. Fix any such $L$. Let $\mu : B \to \mathcal{A}$ satisfy $\mu(y) = G_y$ for all $y \in B$. Since $G_y(R) = y$ for all $y \in B$, $\mu$ is a bijection and $L_i^R$ is a linear order for all $i$. By compatibility, $\mu L = R|_B$. Since $F(L) = F = G_x$, neutrality implies that $F(R|_B) = F(\mu L) = \mu^{-1}(F(L)) = \mu^{-1}(G_x) = x = F(R)$.

CLAIM 2.2. $(\forall m, R, x, y)[((x \neq y) \wedge ((\forall i)[x R_i y])) \Rightarrow (F(R) \neq y)]$.

The proof is by contradiction. Suppose there exist some $m$, $R$, and distinct $x, y \in I_m$ such that $x R_i y$ for all $i$ and $F(R) = y$. Fix any such $m$, $R$, $x$, and $y$. By Claim 2.1, $F(R|_{\{x,y\}}) = y$, but by unanimity, $F(R|_{\{x,y\}}) = x$, which is a contradiction.

CLAIM 2.3. $(\exists k)(\forall m, R, x, y)[((x \neq y) \wedge (\{i \mid x R_i y\} = \{k\})) \Rightarrow (F(R|_{\{x,y\}}) = x)]$.

The proof is direct. Let $m = n$, and let $\tilde{R}$ satisfy, for all $i$ and $r \in I_n$,

$$t_r(\tilde{R}_i) = \begin{cases} i + r - 1 & \text{if } i + r - 1 \leqslant n \\ i + r - 1 - n & \text{if } i + r - 1 > n \end{cases} \tag{15}$$

Let $k = F(\tilde{R})$ and $l = t_n(\tilde{R}_k)$. By Claim 2.1, $F(\tilde{R}|_{\{k,l\}}) = F(\tilde{R}) = k$. Further, $t_1(\tilde{R}_k) = k$ and $l \tilde{R}_i k$ for all $i \in N \setminus \{k\}$. Thus, $\{i \mid k \tilde{R}_i l\} = \{k\}$. Consider any $m$, $R$, and distinct $x, y \in I_m$ such that $\{i \mid x R_i y\} = \{k\}$. Let $\mu : \{x, y\} \to \{k, l\}$ satisfy $\mu(x) = k$ and $\mu(y) = l$. Since $\{i \mid x R_i y\} = \{i \mid k \tilde{R}_i l\}$, it follows that $R|_{\{x,y\}} = \mu(\tilde{R}|_{\{k,l\}})$. Hence, by neutrality, $F(R|_{\{x,y\}}) = \mu^{-1}(F(\tilde{R}|_{\{k,l\}})) = \mu^{-1}(k) = x$.

CLAIM 2.4. $F \in \mathcal{F}_d$.

The proof is direct. Fix $k$ as in Claim 2.3. Consider any $m$ and $R$. Set $x = t_1(R_k)$ and $z = m + 1$. Let $\tilde{R} \in \mathcal{L}(I_{m+1})^N$ satisfy

$$(\tilde{R}|_{I_m} = R) \wedge (t_2(\tilde{R}_k) = z) \wedge ((\forall i \neq k)[t_1(\tilde{R}_i) = z]) \tag{16}$$

Since $z \tilde{R}_i y$ for all $i$ and all $y \in I_m \setminus \{x\}$, Claim 2.2 implies that $F(\tilde{R}) \in \{x, z\}$. Since $\{i \mid x \tilde{R}_i z\} = \{k\}$, Claim 2.3 implies that $F(\tilde{R}|_{\{x,z\}}) = x$. By Claim 2.1, $F(\tilde{R}) = F(\tilde{R}|_{\{x,z\}}) = x$. Since $F(\tilde{R}) \in I_m$, another application of Claim 2.1 gives $F(R) = F(\tilde{R}|_{I_m}) = F(\tilde{R}) = x = d_k(R)$. Hence, $F = d_k \in \mathcal{F}_d$. □

**ACKNOWLEDGMENTS**. *None.* • **ARTIFICIAL INTELLIGENCE**. *I used ChatGPT 6.2 Astra Ultra, Google Deep Review, and Refine.ink to check the proof and improve the structure, language, and exposition of the manuscript. I reviewed the tools' outputs and take full responsibility for the final manuscript.* • **CONFLICT OF INTEREST**. *None.* • **DATA**. *None.* • **FUNDING**. *None.*